\documentclass[superscriptaddress,twocolumn,prl]{revtex4}
\usepackage{amsmath}
\usepackage{amssymb}
\usepackage{graphicx}
\usepackage{hyperref}
\usepackage{amsfonts}
\usepackage{bbm}
\usepackage{datetime}
\usepackage{doi}
\usepackage{color}
\usepackage{xcolor}

\longdate

\newcommand{\beginsupplement}{%
        \setcounter{table}{0}
        \renewcommand{\thetable}{S\arabic{table}}%
        \setcounter{figure}{0}
        \renewcommand{\thefigure}{S\arabic{figure}}%
        \setcounter{equation}{0}
        \renewcommand{\theequation}{S\arabic{equation}}%
      }

\makeatletter
\newcommand*{\rom}[1]{\expandafter\@slowromancap\romannumeral #1@}
\makeatother

\begin{document}
\title{Three-Dimensional Active Defect Loops} 
\author{Jack Binysh}
%\email{J.Binysh@warwick.ac.uk}
\affiliation{Mathematics Institute, Zeeman Building, University of Warwick, Coventry, CV4 7AL, United Kingdom.}
\author{\v{Z}iga Kos}
%\email{ziga.kos@fmf.uni-lj.si}
\affiliation{Faculty of Mathematics and Physics, University of Ljubljana, Jadranska 19, 1000 Ljubljana, Slovenia.}
\author{Simon \v{C}opar}
%\email{simon.copar@fmf.uni-lj.si}
\affiliation{Faculty of Mathematics and Physics, University of Ljubljana, Jadranska 19, 1000 Ljubljana, Slovenia.}
\author{Miha Ravnik}
\email{miha.ravnik@fmf.uni-lj.si}
\affiliation{Faculty of Mathematics and Physics, University of Ljubljana, Jadranska 19, 1000 Ljubljana, Slovenia.}
\affiliation{Jo\v{z}ef Stefan Institute, Jamova 39, 1000 Ljubljana, Slovenia.}
\author{Gareth P.~Alexander}
\email{G.P.Alexander@warwick.ac.uk}
\affiliation{Department of Physics and Centre for Complexity Science, University of Warwick, Coventry, CV4 7AL, United Kingdom.}

\begin{abstract}
We describe the flows and morphological dynamics of topological defect lines and loops in three-dimensional active nematics and show, using theory and numerical modelling, that they are governed by the local profile of the orientational order surrounding the defects. Analysing a continuous span of defect loop profiles, ranging from radial and tangential twist to wedge $\pm 1/2$ profiles, we show that the distinct geometries can drive material flow perpendicular or along the local defect loop segment, whose variation around a closed loop can lead to net loop motion, elongation or compression of shape, or buckling of the loops. We demonstrate a correlation between local curvature and the local orientational profile of the defect loop, indicating dynamic coupling between geometry and topology. To address the general formation of defect loops in three dimensions, we show their creation via bend instability from different initial elastic distortions. 
\end{abstract}
\date{\today}
\maketitle

Active matter is a class of materials in which the individual constituents continually consume energy to generate work or motion, maintaining the system in dynamic, self-organised, non-equilibrium states~\cite{ramaswamy2010,marchetti2013}. Examples derive readily from the study of living systems, ranging from intracellular organisation to swarming bacteria and flocks of birds, but they can equally be realised synthetically in self-propelled colloids or microtubule mixtures. The formalism of active liquid crystals~\cite{doostmohammadi2018} has emphasised the key role of topology in active matter, showing the significance of active topological defects to `turbulence' in bacterial suspensions~\cite{wensink2012}, cell populations~\cite{duclos2017}, cultures~\cite{kawaguchi2017} and tissues~\cite{saw2017}, as well as synthetic active nematics~\cite{sanchez2012,keber2014}. Central to much of the phenomenology in (quasi) two-dimensional systems is that defects with a $+1/2$ profile actively self-propel, while those with $-1/2$ profile do not~\cite{giomi2014}. 

The major focus to date has been on topological defects in two-dimensional active systems~\cite{pismen2013,thampi2014,shankar2018,cortese2018}, including in curved geometries~\cite{keber2014,khoromskaia2017,ellis2018,henkes2018}; however, recently results have begun to emerge also in three dimensions. Three-dimensional active nematics are governed by the presence of topological defect lines and loops, which exhibit complex structure and topology-affected dynamics. These have been studied numerically in thin slabs~\cite{shendruk2018}, establishing the cross-over from two-dimensional behaviour and the significance of twist distortions, and in the confined geometry of a spherical droplet~\cite{copar2019}, showing the significance of defect loops in the turbulent regime. Recent experiments in a bulk extensile nematic~\cite{duclos2019} confirm the importance of defect loops, particularly those of zero topological charge, and provide insights into their structure, formation and dynamics. 

Active nematics are described by a local orientation ${\bf n}$, called the director, which is a unit line field satisfying the symmetry ${\bf n}\sim -{\bf n}$, and active stresses $-\zeta {\bf n} {\bf n}$ along this orientation~\cite{ramaswamy2010}, where $\zeta$ is a phenomenological constant that is positive in extensile materials and negative in contractile ones. Defects are regions where the local orientation of the active nematic is broken due to frustration and in three dimensions are, usually, in the form of lines, which can close into loops. The geometric structure of a defect line is encoded in its profile, the variation of the director field in a perpendicular cross-section. In the two-dimensional $\pm 1/2$ profiles the director lies in this cross-sectional plane, however, in three dimensions a far greater range of geometric profiles are possible~\cite{friedel1969,copar2019,duclos2019,alexander2012,copar2011,machon2016,copar2011PRE,copar2013}. The topological classification of nematic defect loops is also rich, particularly the interplay between topological charge and orientability~\cite{copar2011,alexander2012,machon2016}.

In this Letter, we describe the three-dimensional flows generated by active defect lines and zero topological charge defect loops, applicable to systems such as microtubule-based active nematics~\cite{sanchez2012,duclos2019}. Using analytic calculations and mesoscopic continuum numerical modelling, we show that the non-linear dynamics of three-dimensional active nematic defects can be understood in terms of their local director profile and associated `self-propulsion velocity'. We then give a statistical analysis of the geometries of defect loops in three-dimensional active turbulence in the confinement of a sphere, showing that the most common profiles are of twist type and that $\pm 1/2$ profiles occur preferentially at places of high curvature. Finally, we describe the process of defect loop formation from the uniform state via the fundamental bend instability of extensile nematics. 

\begin{figure*}[t]
\centering
\includegraphics[width=\textwidth]{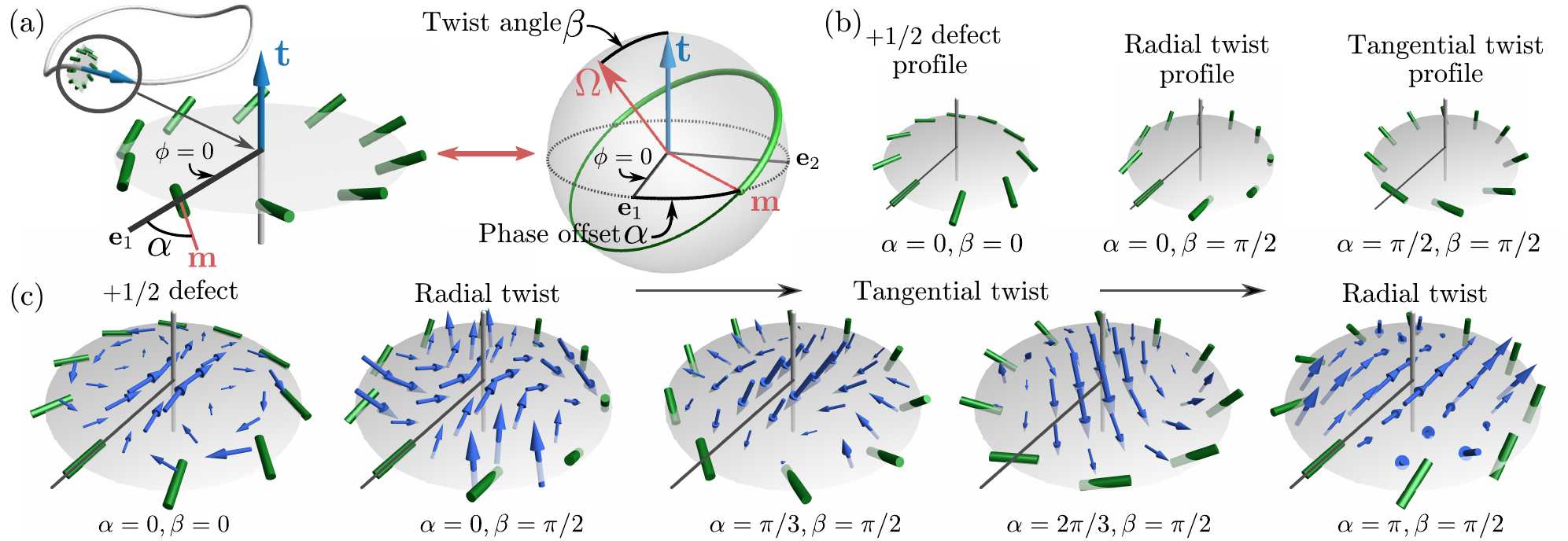}
\caption{Three-dimensional local profiles of defect lines and their flows. (a) A local director profile (green cylinders) corresponds to a path on the unit sphere connecting antipodal points. We consider minimal distortion profiles corresponding to half a great circle (thickened green line) --- these are specified by a twist angle $\beta$ and a phase offset angle $\alpha$. (b) Selected examples of defect profiles of $+1/2$ and `pure twist' type. (c) The active flows (blue arrows) generated by the corresponding nematic defect profiles. We show the case of extensile activity ($\zeta>0$); for contractile activity the direction of the flows reverses.} 
\label{fig:defect_flows}
\end{figure*}

If we plot the variation of the director around any cross-section of a defect line as a path on the unit sphere it will trace out a curve connecting antipodal points, Fig.~\ref{fig:defect_flows}(a). We restrict to those profiles for which this curve is half of a great circle; the rotation of the director around the defect line is then as efficient as possible, which minimises the total elastic distortion. Since any (oriented) great circle can be identified with the vector $\boldsymbol{\Omega}$ normal to it, the space of these minimal distortion profiles is $S^2$; defect loops of this type, described by the rotation vector $\boldsymbol{\Omega}$, were first introduced by Friedel \& de Gennes~\cite{friedel1969}. We parameterise this space as follows: Let ${\bf t}$ denote the unit tangent to the defect line. Any great circle is orthogonal to ${\bf t}$ either at exactly two points, $\pm {\bf m}$, or at every point; the latter are the $\pm 1/2$ profiles of defects in two-dimensional active nematics. We take the defect profile to be the half great circle starting at ${\bf m}$ and ending at $-{\bf m}$ with director   
\begin{equation}
{\bf n} = \cos \frac{1}{2} \phi \,{\bf m} + \sin \frac{1}{2} \phi \Bigl( \cos \beta \,{\bf t} \times {\bf m} + \sin \beta \,{\bf t} \Bigr) ,
\label{eq:geodesic_profile}
\end{equation}
where $\phi$ is the azimuthal angle about the defect line and $\beta$ is the `twist angle' between ${\bf t}$ and $\boldsymbol{\Omega}$. The profile is called `wedge type' ($\pm 1/2$) if $\beta = 0,\pi$ and `twist type' if $\beta = \pi/2$~\cite{duclos2019}; we caution that the word twist here does not hold direct correspondence with a specific type of elastic distortion. We denote by $\alpha$ the `phase offset' between ${\bf m}$ and the direction ${\bf e}_1$ of the radial line $\phi = 0$, and write ${\bf m} = \cos\alpha \,{\bf e}_1 + \sin\alpha \,{\bf e}_2$, where ${\bf e}_2 = {\bf t}\times{\bf e}_1$. Distinct examples of how the phase offset affects the local geometry of twist profiles are given by the cases $\alpha = 0,\pi$ (radial twist) and $\alpha = \pm \pi/2$ (tangential twist), as shown in Fig.~\ref{fig:defect_flows}(b). We remark that this parameterisation is not unique, {\sl e.g.} under the nematic symmetry ${\bf n} \to -{\bf n}$ we have $(\alpha, \beta) \to (\alpha + \pi, -\beta)$.

To determine the nature of the active flows generated by three-dimensional defect lines we adapt the approach of Giomi {\sl et al.}~\cite{giomi2014} for defects in two dimensions. The idea is to solve a Stokes flow problem for an incompressible fluid  
\begin{align}
& - \nabla p + \mu \nabla^2 {\bf u} - \zeta \nabla \cdot \bigl( {\bf n} {\bf n} \bigr) = 0 , && \nabla \cdot {\bf u} = 0 ,
\end{align}
with an active forcing term $-\zeta \nabla \cdot ({\bf n} {\bf n})$ and prescribed director field Eq.~\eqref{eq:geodesic_profile}. The solution is obtained by taking a Helmholtz decomposition of the active force; we present this in the Supplemental Material and show examples in Fig.~\ref{fig:defect_flows}(c), focusing on a family interpolating between profiles of radial and tangential twist. By evaluating the flow at the location of the defect itself we obtain a `self-propulsion velocity' 
\begin{equation}
\begin{split}
{\bf u}_{\textrm{sp}} & = - \frac{\zeta R_{\perp} (1+\cos\beta)^2}{16 \mu} \bigl[ \cos 2\alpha \,{\bf e}_1 + \sin 2\alpha \,{\bf e}_2 \bigr] \\
& \quad - \frac{\zeta R_{\parallel} \sin\beta (1+\cos\beta)}{4\mu} \sin \alpha \,{\bf t} ,
\end{split}
\label{eq:self-propulsion}
\end{equation}
where $R_{\perp}, R_{\parallel}$ are length scales roughly corresponding to the typical defect spacing, or the (local) radius of curvature of the defect loop. More formally, they are constants of integration arising from the choice of boundary conditions in exactly the same way as in the calculation for defects in two dimensions~\cite{giomi2014}. This result has several immediate interesting aspects. First, the magnitude of the transverse velocity depends only on $\beta$ while its direction is determined by ${\bf m}$ and rotates relative to the radial direction $\phi=0$ at the rate $2\alpha$. Second, generically there is also a component of motion (and flow) directed along the defect line; it has maximum strength when $\alpha=\pm \pi/2$ and $\beta = \pm \pi/3$ and vanishes only at profiles with $\alpha=0,\pi$, or $\beta=\pi$. These tangential flows locally stretch the defect line creating elongated sections of loops with such profiles. Third, the self-propulsion velocity vanishes entirely only when $\beta=\pi$, corresponding to the $-1/2$ profile. 

\begin{figure*}[t]
\centering
\includegraphics[width=\textwidth]{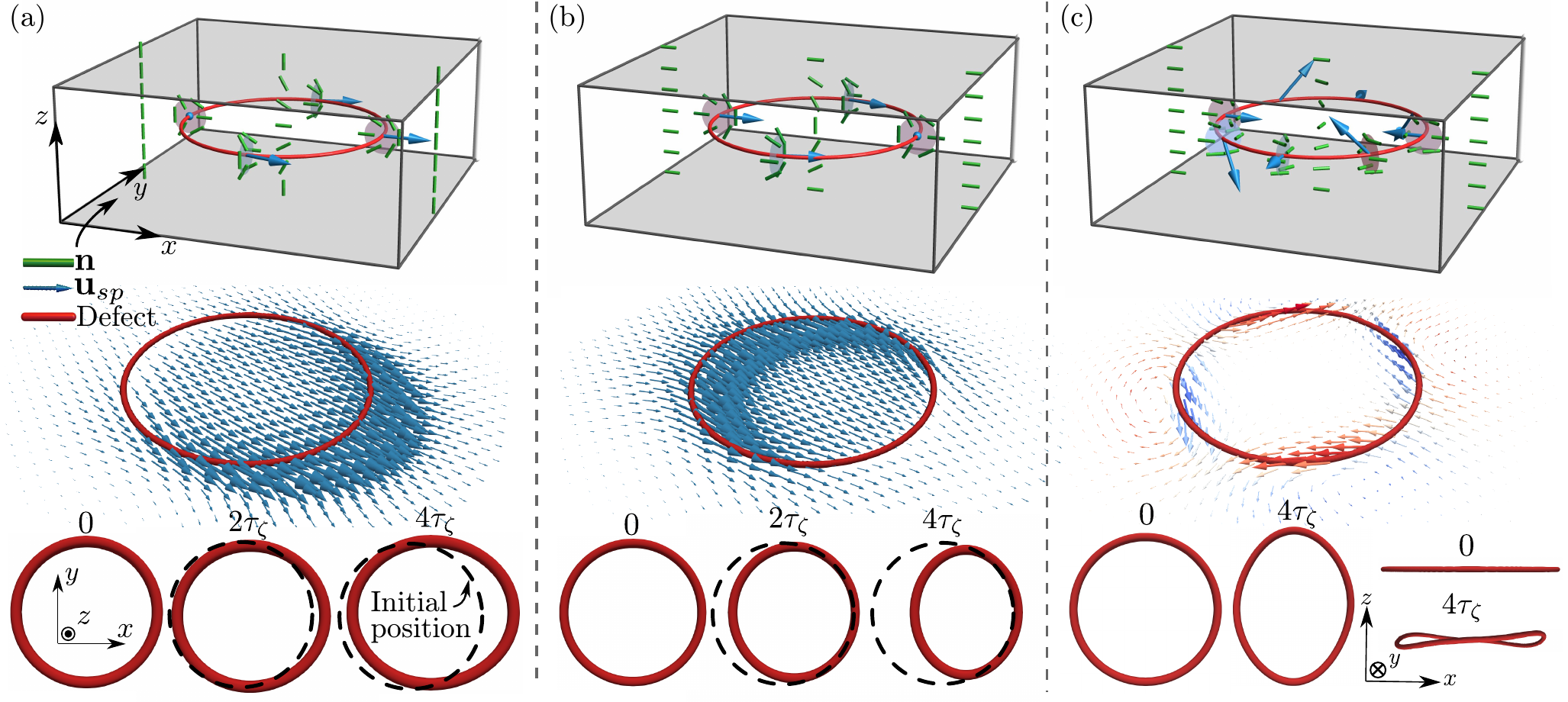}
\caption{Flows and dynamics of different active nematic defect loops with zero topological charge. (a) Defect loop surrounding a cylindrical splay domain (director shown in green). The top panel shows a schematic with the analytically predicted self-propulsion velocity (blue arrows), the middle panel shows the numerically calculated flows in the cell midplane and the bottom panel shows the dynamics of the defect loop position and shape. (b) Defect loop surrounding a cylindrical bend domain with the same comparison between analytic predictions and full simulation as in (a). (c) Defect loop surrounding a cylindrical twist domain. Here the profile is of twist type ($\beta=\pi/2$) everywhere and $\alpha$ makes two full rotations around the loop. The self-propulsion velocities contract the loop along $x$, expand it along $y$ and buckle it on its diagonals. Colour on middle panel velocity gives its $z$ component ($+z$ red, $-z$ blue). Time in all panels is given in units of the active timescale $\tau_\zeta$ (see Supplemental Material).} 
\label{fig:loops}
\end{figure*} 

This calculation qualitatively predicts the behaviour of three-dimensional active defect loops by assigning, at each point, a local profile given by Eq.~\eqref{eq:geodesic_profile} and the corresponding self-propulsion velocity Eq.~\eqref{eq:self-propulsion}. 
In doing this we neglect contributions coming from the curvature of the defect loop but capture the important leading behaviour of self-propulsion. 
Figure~\ref{fig:loops} shows analytical results (top panels) and full numerical simulations (middle and bottom panels) of active flows and dynamics for three distinct types of defect loop with zero topological charge, initiated as surrounding cylindrical domains of either a splay, bend, or twist elastic distortion; these loop types directly correspond to those identified in recent experiments~\cite{duclos2019}. In each case, we consider a slab geometry, confined along $z$ and periodic in $x$, $y$, and initialise a circular defect loop in the cell midplane, working in the regime of weak activity below the threshold for onset of spontaneous flows~\cite{voituriez2005,marenduzzo2007}. The director field around a defect loop with arbitrary geometry (and topology) can be created using Maxwell's solid angle function; as this function is harmonic, this creates director fields that satisfy the (one-elastic-constant) Euler-Lagrange equations~\cite{binysh2018,friedel1969}. Details of numerical methods and parameter values are given in the Supplemental Material. 

We show in Fig.~\ref{fig:loops}(a) a defect loop surrounding a splay domain, with the director uniformly along $z$ outside the loop and matching to normal anchoring at both boundaries. Inside the defect loop the director rotates to point along $x$ in the cell midplane, giving a splay distortion there. A full mapping of the local profiles along the defect loop to Eq.~\eqref{eq:geodesic_profile} is given in the Supplemental Material, but it suffices to focus on only a few points. In the $xz$-plane the local profile is $+1/2$ at positive $x$ and $-1/2$ at negative $x$; the associated active flows cause the $+1/2$ point to propel along positive $x$, while the $-1/2$ point does not intrinsically self-propel. In the $yz$-plane the profile is of tangential twist type with $\beta=\pi/2, \alpha = \pm \pi/2$ and at both points the tangential component of the flow is along positive $x$, stretching these sections. Putting these local pieces together, the defect loop self-propels along positive $x$ and expands into a prolate shape as it does so. This prediction is in remarkably close qualitative agreement with numerical solutions for the velocity field and loop evolution from the full active nematic equations for the velocity. 

Switching to planar alignment along the $x$-direction with the director vertical inside the defect loop generates a loop surrounding a bend domain, shown in Fig.~\ref{fig:loops}(b). Here, in the $xz$-plane, we have a $+1/2$ defect at negative $x$, a $-1/2$ defect at positive $x$, and tangential twist profiles with $\beta = \pi/2, \alpha = \pm \pi/2$ in the $yz$-plane. This leads to the loop self-propelling along positive $x$, this time shrinking and adopting an oblate shape.

The behaviour is different for a defect loop surrounding a (right-handed) twist domain, shown in Fig.~\ref{fig:loops}(c). The director rotates within the $xy$-plane of the cell so that every cross-sectional profile is of pure twist type, with $\beta=\pi/2$ in Eq.~\eqref{eq:geodesic_profile}. For such a defect loop, the transverse component of the self-propulsion velocity has constant magnitude and winds around the defect loop with linking number $-2$ (see Supplemental Material); its direction is such as to compress the loop along the $x$-axis and expand it along $y$, to push it down at points lying along the $x=y$ diagonal and up at points on the $x=-y$ diagonal. The net effect is to cause the defect loop to buckle into a non-planar shape (the sense of buckling reverses upon reversal of the twist domain handedness). 

\begin{figure}[t]
\centering
\includegraphics[width=0.9\columnwidth]{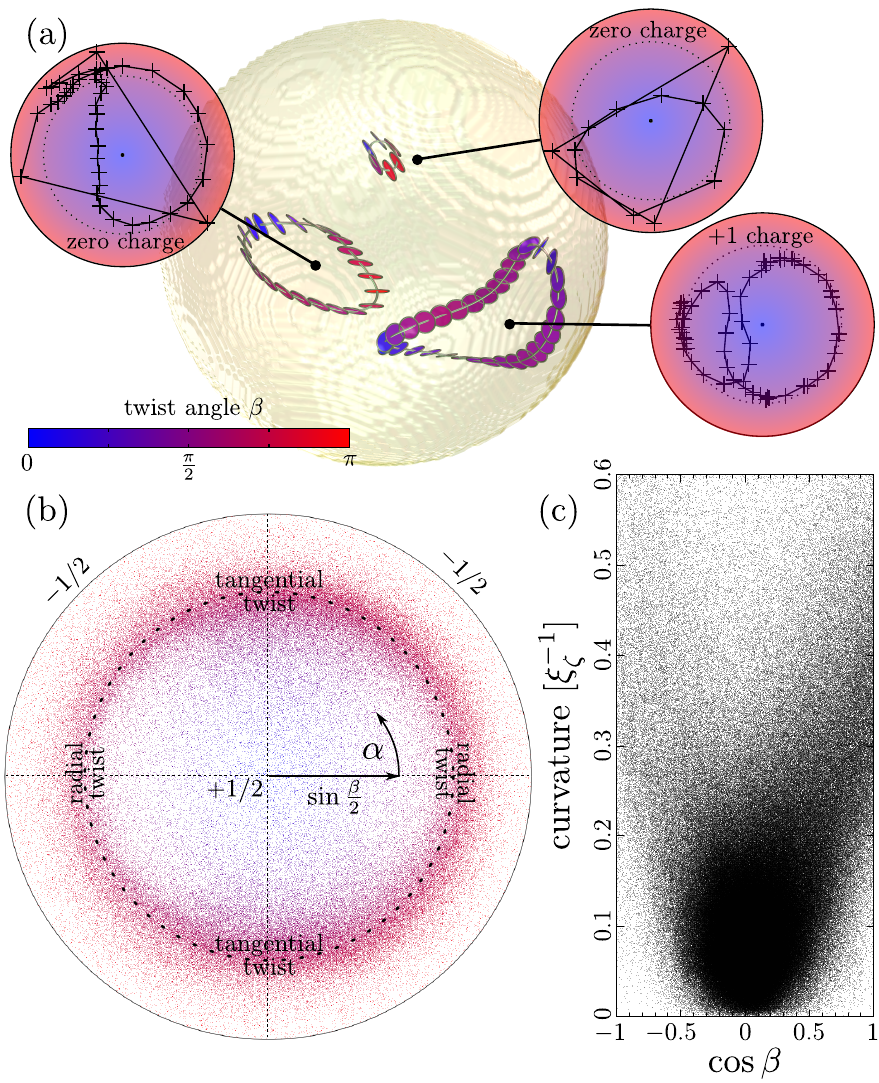}
\caption{Statistics of defect loop profiles. (a) Typical snapshot of a spherical droplet in the turbulent regime, showing three defect loops with cross-sections coloured by twist angle $\beta$. Insets show the variation of profiles for each loop in an equal-area projection of the ($\alpha, \beta$) sphere. (b) Cumulative statistics of defect loop profiles (zero topological charge loops only). The majority are of twist type ($\beta \approx \frac{\pi}{2}$), with a maximum at tangential twist profiles ($\alpha \approx \pm\frac{\pi}{2}$). (c) Correlation between local defect line curvature and twist angle $\beta$ (units are inverse active length scale $\xi_{\zeta}^{-1}$, see Supplemental Material).}
\label{fig:statistics}
\end{figure}

We now investigate what the local profile structure along an active defect loop typically is, using simulations of the turbulent state in a spherical droplet where loops are continually produced, annihilated and rewired~\cite{copar2019}. The local profiles of all defect loops are extracted and tracked dynamically over the simulation time (see Supplemental Material for details of the extraction algorithm). Figure~\ref{fig:statistics}(a) shows a typical simulation snapshot in which there are three defect loops and the profiles of each. Due to the spherical confinement, one of the loops has $+1$ topological charge while the others have zero charge; in the following we consider only the zero topological charge loops. 

In Fig.~\ref{fig:statistics}(b) we plot the distribution of local profiles found over the entire simulation using an equal-area projection of the $(\alpha,\beta)$ sphere (Fig.~\ref{fig:defect_flows}(a)). The distribution shows a bias towards profiles of twist type ($\beta \approx \pi/2$) with tangential twist ($\alpha \approx \pm \pi/2$) more common than radial twist ($\alpha \approx 0, \pi$). We speculate that this bias is caused by the flows about tangential twist profiles locally stretching the defect line, as discussed for Fig.~\ref{fig:loops}(a), increasing the length of the loop with this local profile. We also note a symmetry between $(\alpha,\beta)$ and $(-\alpha,\beta)$, a consequence of invariance under $\bf t \rightarrow -t$ coming from the fact that the defect loops are not naturally oriented, but a slight asymmetry between $(\alpha,\beta)$ and $(\alpha+\pi,\beta)$. This asymmetry is detecting the global structure of flows confined to a droplet, which spontaneously pick an axis and sense of rotation, as detailed in~\cite{copar2019}. Finally, in Fig.~\ref{fig:statistics}(c) we plot the twist angle $\beta$ against the local defect line curvature, which shows that the twist profiles ($\cos\beta \approx 0$) coincide with the places of lowest curvature, while the wedge-type $\pm 1/2$ profiles reside at `apex points' where the curvature is highest. This correlation is consistent with the dynamics of the highly idealised loop encircling a splay domain, shown in Fig.~\ref{fig:loops}(a), and the formation of hairpins under the combination of transverse self-propulsion of $+1/2$ points and line stretching of tangential twist profiles. 

\begin{figure}[t]
\centering
\includegraphics[width=\columnwidth]{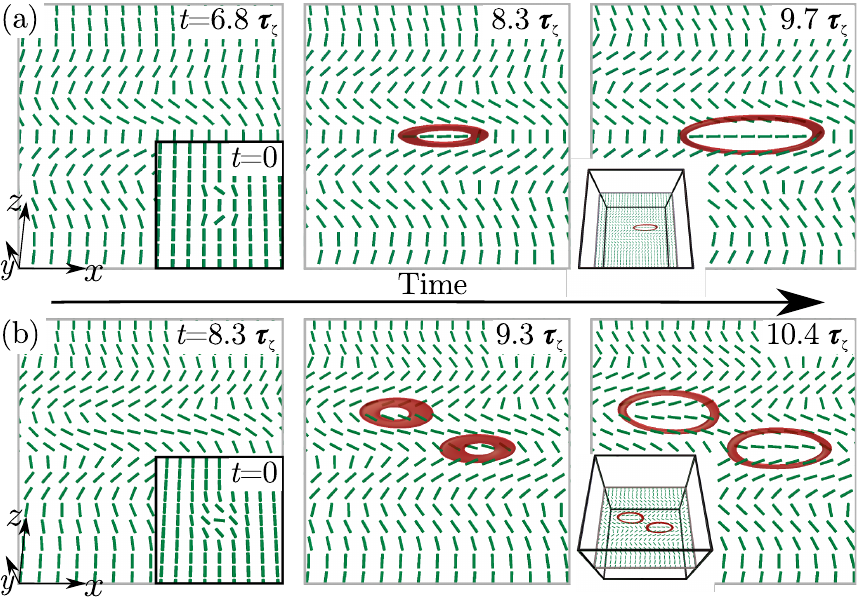}
\caption{Nucleation of zero topological charge active nematic defect loops. (a) A localised bend distortion in the director (inset, green) seeds a defect loop (red) via the generic bend instability. The nucleated loop has the same structure as that surrounding the splay domain shown in Fig.~\ref{fig:loops}(a). (b) Seeds of different elastic distortions generate defect loops of the same type, emerging from regions of maximal bend by the same generic instability.} 
\label{fig:LoopNucleation}
\end{figure}

Defect loops in three-dimensional active nematics can nucleate directly from bend instability, in direct analogy with the nucleation of $\pm 1/2$ point defect pairs in two dimensions~\cite{sanchez2012,thampi2014EPL}. To study this important nucleation process, we initialise a localised bend distortion in a homogeneous far field director, see Fig.~\ref{fig:LoopNucleation}(a). The bend distortion grows until it nucleates a defect loop in the cell midplane, whose structure is exactly that of the defect loop surrounding a splay domain shown in Fig.~\ref{fig:loops}(a). The nucleated loop then self-propels with its $+1/2$ section leading and adopts a prolate shape, with greatest curvature at the points with $\pm 1/2$ profiles. Changing the type of seeded elastic distortion, Fig.~\ref{fig:LoopNucleation}(b), leads to essentially the same nucleation process, with the sites of nucleation being the regions of maximal bend and the defect loops created having the same structure. 

Our results can provide the basis for the analysis and interpretation of three-dimensional active nematic liquid crystals. Especially, it is important to realise that defect lines and loops in 
three-dimensional active nematics can exhibit a full span of different local orientational profiles --- of $\pm 1/2 $ wedge, twist and mixed type --- which results in profoundly different local self-propulsion velocities, both in the directions perpendicular and along the defect loop segment. There are many natural directions for extension of our work, including to a detailed comparison of topologically charged and uncharged loops~\cite{copar2019}, the coupling between loop geometry, biaxiality in defect cores~\cite{krajnik2019} and defects in active cholesterics~\cite{whitfield2017,metselaar2019}. The significance of two-dimensional topological defects to biological systems such as cell cultures and tissues has been well-established in recent years~\cite{duclos2017,kawaguchi2017,saw2017}; active defect loops may provide similar insights to fully three-dimensional biological tissues, fluids and processes.

\acknowledgements{J.B.~was supported by a University of Warwick IAS Early Career Fellowship. S.\v{C}., \v{Z}.K. and  M.R.  acknowledge funding from Slovenian Research Agency (ARRS) under contracts P1-0099, J1-9149, N1-0124 and L1-8135.
M.R. and G.P.A. ~thank the Isaac Newton Institute for Mathematical Sciences for support during the programme ``The mathematical design of new materials'' under EPSRC grant number EP/R014604/1. This work was also supported by a STSM Grant from the COST Action EUTOPIA (number CA17139).}
%J.B.~and G.P.A.~were supported by the UK EPSRC through Grant No. EP/L015374/1. 

\onecolumngrid
\pagebreak
\beginsupplement

\section*{Supplemental Material}

\subsection{I. Active Flows of Three-Dimensional Defect Lines}

We recapitulate the family of three-dimensional defect profiles that we consider. Let ${\bf t}$ be the local tangent vector to a defect line. At each point the director field corresponds to a line element, or a pair of antipodal points $\{{\bf n},-{\bf n}\}$ on the unit sphere $S^2$ of directions in space. The local profile is a variation of ${\bf n}$ that sweeps out a path on $S^2$ connecting antipodal points. The most direct such paths (which have lowest energy in a one-elastic-constant approximation) are geodesics, or halves of great circles. Any such geodesic intersects the equator orthogonal to ${\bf t}$ in a pair of points $\pm {\bf m}$, or is entirely contained in the equator; the latter happens only for the $\pm 1/2$ profiles, which are thus seen to be exceptional. We choose the orientation ${\bf m}$ to `frame' the local profile, {\sl i.e.} the director points along ${\bf m}$ when $\phi=0$. Then a general (geodesic) local profile can be written 
\begin{equation}
{\bf n} = \cos \frac{1}{2} \phi \,{\bf m} + \sin \frac{1}{2} \phi \Bigl( \cos \beta \,{\bf t} \times {\bf m} + \sin \beta \,{\bf t} \Bigr) ,
\label{eqS:geodesic_family}
\end{equation}
where $\beta$ is the `twist angle' between the tangent vector ${\bf t}$ and the normal $\boldsymbol{\Omega}$ to the great circle corresponding to the local profile. We denote the direction of the radial line $\phi=0$ by ${\bf e}_1$ and set ${\bf e}_2={\bf t} \times {\bf e}_1$. Then the orientation ${\bf m}$ along $\phi=0$ can be written ${\bf m}=\cos\alpha \,{\bf e}_1 + \sin\alpha \,{\bf e}_2$. The angles $(\alpha,\beta)$ give a parameterisation of the family of local profiles we consider. This parameterisation is not unique as, for instance, under the nematic symmetry ${\bf n} \to -{\bf n}$ we have $(\alpha,\beta) \to (\alpha+\pi,-\beta)$. When $\beta = 0,\pi$ the profile is termed `wedge type' ($\pm 1/2$), while when $\beta = \pi/2$ it is termed `twist type'~\cite{duclos2019S}. We call $\alpha$ the `phase offset angle': a twist profile ($\beta = \pi/2$) with $\alpha = 0,\pi$ has director oriented radially along the line $\phi = 0$, which we call a `radial twist profile', and one with $\alpha = \pm\pi/2$ has director aligned in the tangential direction along the same $\phi = 0$ line, which we call a `tangential twist profile'. These geometric differences in the local director profile lead to different active flows and associated dynamics. 

To determine the active flows generated by three-dimensional defect lines we adapt the approach of Giomi {\sl et al.}~\cite{giomi2014S} for the $\pm 1/2$ defects in two dimensions. Their approach consists of solving the Stokes equation for an incompressible fluid with active forcing
\begin{align}
& - \nabla p + \mu \nabla^2 {\bf u} = \zeta \nabla \cdot \bigl( {\bf n} {\bf n} \bigr) , && \nabla \cdot {\bf u} = 0 ,
\label{eqS:Stokes}
\end{align} 
and specified director field corresponding to the defect profile~\eqref{eqS:geodesic_family}. A convenient method for developing the solution is to use a Helmholtz decomposition to write the active force as a sum of a curl-free gradient term and a divergence-free curl term; the former is balanced by a pressure and the latter generates fluid flows~\cite{green2017S}. 

Following a short calculation the active force is determined to be 
\begin{equation}
\begin{split}
\nabla \cdot \bigl( {\bf n} {\bf n} \bigr) & = \nabla \biggl( \frac{\sin^2 \beta}{4} \cos \phi \biggr) + \frac{(1+\cos \beta)^2}{8r} \bigl[ \cos 2\alpha \,{\bf e}_1 + \sin 2\alpha \,{\bf e}_2 \bigr] \\
& \quad - \frac{(1-\cos\beta)^2}{8r} \Bigl[ \cos 2\alpha \bigl( \cos 2\phi \,{\bf e}_1 - \sin 2\phi \,{\bf e}_2 \bigr) + \sin 2\alpha \bigl( \sin 2\phi \,{\bf e}_1 + \cos 2\phi \,{\bf e}_2 \bigr) \Bigr] \\ 
& \quad + \frac{\sin\beta}{4r} \Bigl[ \bigl( 1 + \cos \beta \bigr) \sin \alpha + \bigl( 1 - \cos \beta \bigr) \bigl( \sin \alpha \cos 2\phi - \cos \alpha \sin 2\phi \bigr) \Bigr] {\bf t} .
\end{split}
\end{equation}
Performing the Helmholtz decomposition we then find that the pressure is given by 
\begin{equation}
p = - \frac{\zeta \sin^2 \beta}{4} \cos \phi - \frac{\zeta (1+\cos\beta)^2}{8} \Bigl[ \cos 2\alpha \cos \phi + \sin 2\alpha \sin \phi \Bigr] + \frac{\zeta (1-\cos\beta)^2}{24} \Bigl[ \cos 2\alpha \cos 3\phi + \sin 2\alpha \sin 3\phi \Bigr] ,
\end{equation}
and the fluid velocity is 
\begin{equation}
\begin{split}
{\bf u} & = \frac{\zeta (1+\cos\beta)^2}{16 \mu} \biggl( (r-R_{\perp}) \Bigl[ \cos 2\alpha \,{\bf e}_1 + \sin 2\alpha \,{\bf e}_2 \Bigr] \\
& \qquad \qquad - \frac{r}{3} \Bigl[ \cos 2\alpha \bigl( \cos 2\phi \,{\bf e}_1 + \sin 2\phi \,{\bf e}_2 \bigr) + \sin 2\alpha \bigl( \sin 2\phi \,{\bf e}_1 - \cos 2\phi \,{\bf e}_2  \bigr) \Bigr] \biggr) \\
& \quad + \frac{\zeta (1-\cos\beta)^2}{48 \mu} \, r \biggl( \cos 2\alpha \Bigl[ \Bigl( \cos 2\phi + \frac{1}{5} \cos 4\phi \Bigr) {\bf e}_1 - \Bigl( \sin 2\phi - \frac{1}{5} \sin 4\phi \Bigr) {\bf e}_2 \Bigr] \\
& \qquad \qquad \qquad + \sin 2\alpha \Bigl[ \Bigl( \sin 2\phi + \frac{1}{5} \sin 4\phi \Bigr) {\bf e}_1 + \Bigl( \cos 2\phi - \frac{1}{5} \cos 4\phi \Bigr) {\bf e}_2 \Bigr]  \biggr) \\
& \quad + \frac{\zeta \sin\beta}{4\mu} \biggl( \sin \alpha \bigl( 1 + \cos \beta \bigr) (r-R_{\parallel}) - \bigl( 1 - \cos \beta \bigr) \frac{r}{3} \Bigl[ \sin \alpha \cos 2\phi - \cos \alpha \sin 2\phi \Bigr] \biggr) {\bf t} ,
\end{split}
\end{equation}
where $r$ is the radial distance to the defect line and $R_{\perp}, R_{\parallel}$ are constants of integration (associated to solutions of the homogeneous problem), which correspond physically to a `self-propulsion velocity'. We remark that the solution for the $-1/2$ profile obtained when $\beta=\pi$ (and $\alpha=0$) differs from the one given by Giomi {\sl et al.}~\cite[(3.9b)]{giomi2014S} in the sign for $u_2$, and also by a solution of the homogeneous problem; the former corrects a typographical error in their paper. 

Finally, evaluating the flow field at $r=0$ we obtain the self-propulsion velocity of the local profile 
\begin{equation}
{\bf u}_{\textrm{sp}} = - \frac{\zeta R_{\perp} (1+\cos\beta)^2}{16 \mu} \Bigl[ \cos 2\alpha \,{\bf e}_1 + \sin 2\alpha \,{\bf e}_2 \Bigr] - \frac{\zeta R_{\parallel} \sin\beta (1+\cos\beta)}{4\mu} \sin \alpha \,{\bf t} ,
\end{equation}
as stated in the main text. 

\subsubsection{Elastic Distortions of Three-Dimensional Defect Line Profiles}

The terminology `twist type' for profiles with $\beta = \pi/2$ has the slight disadvantage that there is potential to conflate the term `twist' with the type of elastic distortion that the profile exhibits. To emphasise that the two uses of the same word should be considered separate, we record here the expressions for the splay ${\bf n}(\nabla\cdot{\bf n})$, twist ${\bf n} \cdot \nabla \times {\bf n}$ and bend $({\bf n}\cdot\nabla){\bf n}$ of the director field~\eqref{eqS:geodesic_family}: 
\begin{align}
\begin{split}
{\bf n} \bigl( \nabla \cdot {\bf n} \bigr) & = \frac{1}{4r} \Bigl[ \sin \phi \sin(\phi - \alpha) + \bigl( 1 + \cos \phi \bigr) \cos(\phi - \alpha) \cos \beta \Bigr] {\bf m} \\
& \quad + \frac{1}{4r} \Bigl[ \bigl( 1 - \cos \phi \bigr) \sin(\phi - \alpha) + \sin \phi \cos(\phi - \alpha) \cos \beta \Bigr] \Bigl( \cos \beta \,{\bf t}\times{\bf m} + \sin \beta \,{\bf t} \Bigr) , 
\end{split} \\
{\bf n} \cdot \nabla \times {\bf n} & = \frac{1}{2r} \cos(\phi - \alpha) \sin \beta , \\
\begin{split}
\bigl( {\bf n} \cdot \nabla \bigr) {\bf n} & = \frac{1}{4r} \Bigl[ \sin \phi \sin(\phi - \alpha) - \bigl( 1 - \cos \phi \bigr) \cos(\phi - \alpha) \cos \beta \Bigr] {\bf m} \\
& \quad + \frac{1}{4r} \Bigl[ - \bigl( 1 + \cos \phi \bigr) \sin(\phi - \alpha) + \sin \phi \cos(\phi - \alpha) \cos \beta \Bigr] \Bigl( \cos \beta \,{\bf t}\times{\bf m} + \sin \beta \,{\bf t} \Bigr) . 
\end{split}
\end{align}
It can be seen that, for the pure twist profiles with $\beta = \pi/2$, the twist in the director field, ${\bf n}\cdot\nabla\times{\bf n}$, neither has a single sign (corresponding to a well-defined handedness) around the defect line nor holds greater significance than the splay and bend distortions. Although it may be felt that a different terminology could help remove potential confusion, we do not propose any such here; it may also be felt that the different uses of `twist' are apparent once the terminology has been stated clearly. 

\subsection{II. Defect Loops and their Local Profiles}

We first review briefly a canonical construction for defect loops in a nematic liquid crystal~\cite{binysh2018S}. Suppose we have a defect loop $K$ in a nematic with uniform far field corresponding to the direction ${\bf d}_1$ and let ${\bf d}_2$ be any unit vector orthogonal to this direction, ${\bf d}_1 \cdot {\bf d}_2 = 0$. Then the director field 
\begin{equation}
{\bf n} = \cos \frac{1}{4} \omega_{K} \,{\bf d}_1 + \sin \frac{1}{4} \omega_K \,{\bf d}_2 ,
\label{eqS:director_loop}
\end{equation}
where $\omega_K$ is the solid angle function for $K$, contains $K$ as a defect loop and matches the far field boundary conditions~\cite{binysh2018S}. At any point ${\bf x}$ the solid angle of $K$ is the area swept out by the projection of $K$ onto the unit sphere centred on ${\bf x}$. A natural geometric formula for it is~\cite{binysh2018S} 
\begin{equation}
\omega_K({\bf x}) = 2\pi \bigl( 1 + \textrm{Wr}_K \bigr) - \int_{K} \frac{{\bf p}\cdot({\bf t}\times d{\bf t})}{1+{\bf p}\cdot{\bf t}} \quad \textrm{mod } 4\pi ,
\end{equation} 
where ${\bf p} = \frac{{\bf y}-{\bf x}}{|{\bf y}-{\bf x}|}$, with ${\bf y}$ a point of $K$, is the direction vector to $K$ from ${\bf x}$, ${\bf t}$ is the unit tangent vector to $K$, and $\textrm{Wr}_K$ is its writhe. 

The family of defect loop director fields~\eqref{eqS:director_loop} contains those described in recent experimental work on three-dimensional active nematics~\cite{duclos2019S} and previously in the literature on passive liquid crystals~\cite{friedel1969S}, with the rotation vector $\boldsymbol{\Omega}$ given by $\boldsymbol{\Omega} = {\bf d}_1 \times {\bf d}_2$. Part of the significance of these director fields is that, since the solid angle is a harmonic function, they satisfy the one-elastic-constant Euler-Lagrange equations~\cite{binysh2018S,friedel1969S} (for the orientational degrees of freedom), for any choice of $K$. Finally, we mention that $K$ is not constrained to be a simple circular loop (as we focus on here) and can have any shape, or be any type of knot or even link. Further details on the construction, including extensions that enable all topological classes of director field to be realised, can be found in~\cite{binysh2018S,alexander2018S}. 

We now explain in detail how to map the the director field~\eqref{eqS:director_loop} onto the family of local profiles~\eqref{eqS:geodesic_family} for the particular cases of circular defect loops that we consider and present in the main text. For concreteness, we take $K$ to be the circle $(R_K \cos \theta, R_K \sin \theta , 0)$ of radius $R_K$ in the $xy$-plane, parameterised by the angle $\theta$ of a global cylindrical coordinate system $(\rho, \theta, z)$. Since we will also refer to a (distinct) local cylindrical coordinate system adapted to the tubular neighbourhood of $K$, we emphasise that these are global cylindrical coordinates defined by $x=\rho \cos \theta$, $y = \rho \sin \theta$. The tangent vector to the defect loop is ${\bf t} = {\bf e}_{\theta}$. The local cylindrical coordinates around any point of $K$ will be denoted $(r,\phi,s)$, with, as in the main text, the unit vector ${\bf e}_1$ corresponding to the direction $\phi = 0$ and ${\bf e}_2 = {\bf t} \times {\bf e}_1$. The direction $\phi = 0$ is defined by the property that the director is orthogonal to the defect line tangent along this direction, which gives a canonical framing to the defect loop. 

\subsubsection{Case 1: Splay Wall}

The far field orientation is ${\bf d}_1 = {\bf e}_z$ (matching normal anchoring on the surfaces of the cell) and we take ${\bf d}_2 = {\bf e}_x$. The director is then orthogonal to the tangent vector ${\bf t}$ where $\omega_K = 0$ mod $4\pi$ -- `outside' the loop --, which gives $\phi = \omega_K /2$, ${\bf e}_1 = {\bf e}_{\rho}$, and consequently ${\bf e}_2 = -{\bf e}_z$. Further, we have ${\bf m} = {\bf e}_z$ everywhere around the loop. Matching these identifications onto~\eqref{eqS:geodesic_family} gives $\alpha = -\pi/2$ and $\beta = \theta$. The self-propulsion velocity, expressed in the global cylindrical coordinate system, is 
\begin{equation}
{\bf u}_{\textrm{sp}} = \frac{\zeta R_{\perp} (1+\cos\theta)^2}{16 \mu} {\bf e}_{\rho} + \frac{\zeta R_{\parallel} \sin\theta (1+\cos\theta)}{4\mu} \,{\bf e}_{\theta} .
\end{equation}

\subsubsection{Case 2: Bend Wall}

The far field orientation is ${\bf d}_1 = {\bf e}_x$ (matching tangential anchoring on the surfaces of the cell) and we take ${\bf d}_2 = {\bf e}_z$. The director is orthogonal to the tangent vector ${\bf t}$ where $\omega_K = 2\pi$ mod $4\pi$ -- `inside' the loop --, which gives $\phi = \omega_K /2 - \pi$, ${\bf e}_1 = -{\bf e}_{\rho}$ and ${\bf e}_2 = {\bf e}_z$. We again have ${\bf m} = {\bf e}_z$ everywhere around the loop, and matching onto~\eqref{eqS:geodesic_family} gives $\alpha = \pi/2$ and $\beta = \pi - \theta$. The self-propulsion velocity, expressed in the global cylindrical coordinate system, is 
\begin{equation}
{\bf u}_{\textrm{sp}} = - \frac{\zeta R_{\perp} (1-\cos\theta)^2}{16 \mu} {\bf e}_{\rho} - \frac{\zeta R_{\parallel} \sin\theta (1-\cos\theta)}{4\mu} \,{\bf e}_{\theta} .
\end{equation} 

\subsubsection{Case 3: Twist Wall}

The far field orientation is ${\bf d}_1 = {\bf e}_x$ (matching tangential anchoring on the surfaces of the cell) and we take ${\bf d}_2 = {\bf e}_y$ for a right-handed twist wall. Expressing the director in the global cylindrical basis we obtain ${\bf n} = \cos \bigl(\frac{1}{4} \omega_K - \theta \bigr) \,{\bf e}_{\rho} + \sin \bigl(\frac{1}{4} \omega_K - \theta \bigr) \,{\bf e}_{\theta}$. The director is orthogonal to the tangent vector ${\bf t}={\bf e}_{\theta}$ along the locus $\omega_K = 4\theta$ mod $4\pi$, which defines a ribbon that has linking number $+2$ with the defect line. We have $\phi = \omega_K /2 - 2\theta$ and the direction vector ${\bf e}_1$ is $\cos 2\theta \,{\bf e}_{\rho} - \sin 2\theta \,{\bf e}_z$. Matching these identifications onto~\eqref{eqS:geodesic_family} gives $\alpha = -2\theta$ and $\beta = \pi/2$. The self-propulsion velocity, expressed in the global cylindrical coordinate system, is 
\begin{equation}
{\bf u}_{\textrm{sp}} = - \frac{\zeta R_{\perp}}{16 \mu} \bigl[ \cos 2\theta \,{\bf e}_{\rho} + \sin 2\theta \,{\bf e}_z \bigr] + \frac{\zeta R_{\parallel}}{4\mu} \sin 2\theta \,{\bf e}_{\theta} .
\end{equation}
The transverse flows are of constant magnitude, and wind around the defect loop with linking number $-2$, becoming planar and flowing inwards (outwards) at $\theta = 0, \pi$ ($\theta = \pi/2,-\pi/2$). At $\theta = \pi/4, 3\pi/4$ {\sl etc.}\ they are binormal and alternate in sign, causing the observed buckling. 

For a left-handed twist wall we may simply make the replacement ${\bf d}_2 \to -{\bf e}_y$. Repeating the preceding analysis we find $\alpha = 2\theta$ and $\beta = -\pi/2$. The self-propulsion vecocity is 
\begin{equation}
{\bf u}_{\textrm{sp}} = - \frac{\zeta R_{\perp}}{16 \mu} \bigl[ \cos 2\theta \,{\bf e}_{\rho} - \sin 2\theta \,{\bf e}_z \bigr] + \frac{\zeta R_{\parallel}}{4\mu} \sin 2\theta \,{\bf e}_{\theta} ,
\end{equation}
giving an opposite sense of buckling, but identical radial and tangential flows, to the right-handed twist wall. 

\subsection{III. Details of Numerical Simulation}

We simulate active liquid crystals numerically using a hybrid lattice Boltzmann algorithm to solve the Beris-Edwards equations for passive liquid crystals augmented by an active contribution to the stress tensor~\cite{marenduzzo2007S,doostmohammadi2018S,ramaswamy2010S,BerisEdwardsS}. These are a set of coupled non-linear equations for the Q-tensor order parameter of the liquid crystal and the fluid velocity. The order and elasticity of the liquid crystal is set by the Landau-de Gennes free energy 
\begin{equation}
F = \int \biggl\{ \frac{L_1}{2} \Bigl(\partial_k Q_{lj}\Bigr)^2  + \frac{L_2}{2} \Bigl( \partial_j Q_{ij} \Bigr)^2 + \frac{A}{2} \, Q_{ij} Q_{ij} + \frac{B}{3} \, Q_{ij} Q_{ik} Q_{jk} + \frac{C}{4} \bigl( Q_{ij} Q_{ij} \bigr)^2 \biggr\} dV ,
\end{equation}
where $L_1, L_2$ are elastic moduli and $A, B, C$ are bulk material coefficients, whose values are chosen so as to be in the ordered phase. Here we work with a one-elastic-constant nematic, corresponding to $L_1= L$ and $L_2=0$. The Q-tensor evolves dynamically according to the advective relaxational equation 
\begin{equation}
\bigl( \partial_t + u_k \partial_k \bigr) Q_{ij} + \Omega_{ik} Q_{jk} + Q_{ik} \Omega_{jk} = \Gamma H_{ij} + \lambda D_{ik} \Bigl( Q_{jk} + \frac{1}{3} \delta_{jk} \Bigr) + \lambda \Bigl( Q_{ik} + \frac{1}{3} \delta_{ik} \Bigr) D_{jk} - 2 \lambda \Bigl( Q_{ij} + \frac{1}{3} \delta_{ij} \Bigr) Q_{kl} D_{kl}  ,
\end{equation}
where $H_{ij} = - \delta F / \delta Q_{ij} + \frac{1}{3} \delta_{ij} \text{Tr}(\delta F / \delta Q_{kl})$ is the molecular field, $\Gamma$ is a relaxational material constant, $\lambda$ is a flow alignment parameter, and $D_{ij} = \frac{1}{2} (\partial_i u_j + \partial_j u_i)$ and $\Omega_{ij} = \frac{1}{2} (\partial_i u_j - \partial_j u_i)$ are the symmetric and antisymmetric parts of the velocity gradients. The fluid velocity satisfies the continuity, $\partial_t \rho + \partial_i (\rho u_i) = 0$, and Navier-Stokes, $\rho (\partial_t + u_j \partial_j) u_i = \partial_j \sigma_{ij}$, equations, with the stress tensor  
\begin{equation}
\begin{split}
\sigma_{ij} & = - p\, \delta_{ij} + 2\mu D_{ij} + 2\lambda \Bigl( Q_{ij} + \frac{1}{3} \delta_{ij} \Bigr) Q_{kl} H_{kl} - \lambda H_{ik} \Bigl( Q_{jk} + \frac{1}{3} \delta_{jk} \Bigr) - \lambda \Bigl( Q_{ik} + \frac{1}{3} \delta_{ik} \Bigr) H_{jk}  \\ 
& \quad + Q_{ik} H_{jk} - H_{ik} Q_{jk} - \partial_{i} Q_{kl} \, \frac{\delta F}{\delta \partial_j Q_{kl}} - \zeta Q_{ij} ,
\end{split}
\end{equation}
where $\rho$ is the density, $p$ is the pressure and $\mu$ is a Newtonian viscosity. The activity is provided by the final term, $-\zeta Q_{ij}$, which corresponds to a force dipole at the microscopic level, with $\zeta$ a phenomenological parameter that is positive in extensile materials and negative in contractile ones~\cite{ramaswamy2010S}. For visualisation it is convenient to represent the order in terms of the liquid crystal director $n_i$, which is most commonly expressed as related to the Q-tensor through the uniaxial form $Q_{ij} = \frac{3s}{2} (n_i n_j - \frac{1}{3} \delta_{ij})$, where $s$ is the scalar magnitude of the order parameter. Numerically we obtain $n_i$ from $Q_{ij}$ as the eigenvector associated to the largest eigenvalue. 

Fundamental length and time scales in this system are given by the nematic correlation length $\xi_n = \sqrt {L/\left(A+Bs_{\textrm{eq}}+\frac{9}{2}C s_{\textrm{eq}}^2\right)}$, where $s_{\textrm{eq}}$ is the equilibrium value of the scalar order parameter, and the nematic timescale $\tau_n = \xi_n^2/\Gamma L$; $\xi_{n}$ determines the typical defect line core size. A second set of scales for active phenomena are the active length and time scales, $\xi_{\zeta}=\sqrt{L/\zeta}$ and $\tau_{\zeta}=\mu/\zeta$. We may compare $\xi_\zeta$ to a confining lengthscale $D$ with the dimensionless activity number $Ac = D/\xi_\zeta = \sqrt{\zeta D^2/L}$. In Table \ref{table:parameters} we list the parameter valued used here, expressed in units of the basic nematic scales $L, \xi_n, \tau_n$; identical choices are made in the recent work \cite{copar2019S}. Simulations in Fig.~2 were performed at $\zeta = 0.044 L/\xi_n^2$, giving an active lengthscale of $\xi_\zeta = 4.8 \xi_n$ and active timescale of $\tau_\zeta = 31 \tau_n$. The confining lengthscale in our slab geometry is $D=106\xi_n$, corresponding to a dimensionless activity number of $Ac = 22$. The following activity values were used in other figures: $\zeta = 0.04 L/\xi_n^2$ in Fig.~3(a); $\zeta = 0.054 L/\xi_n^2$ in Fig.~3(b,c); and $\zeta = 0.3 L/\xi_n^2$ in Fig.~4. Simulations in Fig.~3 were performed in a droplet with radius $136 \xi_n$. Simulations in Fig.~4 were performed in a rectangular simulation box of size $(100 \xi_n)^3$ with periodic boundary conditions. 

\begin{table}
  \caption{Summary of the parameter values used in numerical simulations.}
    \begin{tabular}{  c  c }
    \hline
    Parameter & Value  \\ \hline
    \multicolumn{2}{c}{Landau-de Gennes} \\
    $A$ &$-0.190 L/\xi_n^2$ \\  
    $B$ & $-2.36L/\xi_n^2$ \\  
    $C$ & $1.92L/\xi_n^2$\\  
    $s_{\textrm{eq}}$ & $0.53$\\  
    $\Gamma$ & $\xi_{n}^2/\tau_n L$\\  
    \multicolumn{2}{c}{Beris-Edwards} \\
    $\lambda$ & $1$\\  
    $\mu$ & $1.38 L\tau_n/\xi_n^2$\\  
      $\rho$ & $6.9\cdot10^{-2} \tau_n L/ \xi_{n}^2$ \\  
    \multicolumn{2}{c}{Gridspacing, Timestep} \\
    $\Delta x$ & $1.5\xi_n$\\  
      $\Delta t$ & $0.056\tau_n$\\  
    \hline
    \end{tabular}
  \label{table:parameters}
\end{table}

\subsection{IV. Statistics of Defect Loop Profiles in Turbulent Active Droplets}

Simulation of active nematic turbulence in a droplet with homeotropic surface anchoring was performed for a droplet with radius $135\xi_n$ over simulation time $880 \tau_{\zeta}$ in addition to the initial equilibration time. The time frame of the simulation spans enough dynamics and topological events -- break-ups, annihilations -- to faithfully represent a long-term average.
To retrieve the full director profile information from the simulation, disclination loops were first traced by finding connected lines of low scalar order parameter \cite{sec2014S,copar2019S}. With the loop geometry extracted in the form of a naturally parameterised curve ${\bf r}(s)$, a right-handed coordinate frame was attached to it, defining a local triple $\{ {\bf t}, {\bf u}, {\bf v} \}$ of the tangent and two normal directions. The frames were adjusted to close up continuously with zero self-linking number, to ensure a fixed topology of the framing.

A second coordinate frame was constructed from the director field in close proximity $\epsilon$ of the disclination loop:
\begin{align}
{\bf n}_{+} & = {\bf n}({\bf r}+\epsilon {\bf u}) , \\
{\bf n}_{-} & = {\bf n}({\bf r}-\epsilon {\bf u}) , \\
\boldsymbol{\Omega} & = {\bf n}_{+} \times {\bf n}_{-} ,
\end{align}
where ${\bf n}_{+}$ and ${\bf n}_{-}$ describe the plane of the great half-circle in the director space. Due to the sign ambiguity of the director, this frame has to be regularised. ${\bf n}_{-}$ was oriented such that its sign was consistent with $\tfrac{\pi}{2}$ rotation of ${\bf n}_{+}$ on the counter-clockwise path around the disclination, and orthogonalised. This was achieved by inspecting the director ${\bf n}({\bf r}+\epsilon {\bf v})$ -- halfway around the semi-circle between the samples ${\bf n}_{+}$ and ${\bf n}_{-}$ in the counter-clockwise direction.

An orthonormal coordinate frame is fully described by a pure rotation that maps the standard basis vectors to the basis of the target frame. In principle, we could use $3\times 3$ matrices with basis vectors as rows. However, $SU(2)$ structure is required to extract the full topological information of the loop, including its topological charge \cite{copar2014S}, so we use the unit quaternion representation of rotations, with basis quaternions ${\bf i}, {\bf j}, {\bf k}$ corresponding to the standard basis vectors (equivalently, one could use Pauli matrices). Flipping the sign of a quaternion does not change the frame obtained by the rotation, but matters when continuous variation of the frame is considered; the signs were chosen to obtain continuously varying unit quaternions $q_{\textrm{frame}}$ and $q_{\textrm{dir}}$ from framings $\{ {\bf t}, {\bf u}, {\bf v} \}$ and $\{ {\bf n}_+, {\bf n}_-, \boldsymbol{\Omega} \}$, respectively.

The topological charge of the loop and the local director profiles both rely on the relative orientation of the profile to the disclination line cross section, $q_{\textrm{rel}}=q_{\textrm{frame}}^{-1}q_{\textrm{dir}}$. This relative rotation now parameterises the local director profile by rotating the $+1/2$ profile 
\begin{equation}
  {\bf n}(\phi) = q_{\textrm{rel}} \biggl( {\bf i}\cos\frac{\phi}{2} + {\bf j}\sin\frac{\phi}{2} \biggr) q^{-1}_{\textrm{rel}} .
  \label{eqS:n_phi}
\end{equation}

The unit quaternions still live on $S^3$. As we are interested in the geometric nature of the profile -- to extract the angles $(\alpha,\beta)$ --, rotation of the profile around the tangent is neglected. A counter-clockwise rotation of the profile by $2\delta$ requires two operations: an active rotation of the director at each point by $2\delta$, and a shift in parameterisation (a passive rotation), $\phi \mapsto \phi - 2\delta$. In expression \eqref{eqS:n_phi}, the latter is equivalent to applying a rotation by $-\delta$ directly to the $+1/2$ profile, resulting in total transformation
\begin{equation}
  q_{\textrm{rel}} \mapsto p^2 q_{\textrm{rel}} p^{-1} , \qquad p = \cos\frac{\delta}{2} + {\bf k} \sin\frac{\delta}{2} .
\end{equation}
To reduce the dimensionality of the parameter space from $S^3$ to $S^2$, we lock $\phi=0$ such that the \emph{rotation axis} lies in the plane perpendicular to the loop tangent, which is achieved when the vector component of the quaternion (the axis of rotation) has zero ${\bf k}$-component. This differs from the parameterisation in Eq.~\ref{eqS:geodesic_family} by having the director lie in-plane at $\phi=2\alpha$ instead of $\phi=0$, but conveniently, this makes the direction of the rotation axis (the vector part of the quaternion) coincide in meaning with previously defined vector ${\bf m}$. The function that projects the quaternion $q_{\textrm{rel}} = w + x{\bf i} + y{\bf j} + z {\bf k}$ down to $S^2$, is 
\begin{equation}
\begin{bmatrix} w \\ x \\ y \\ z \end{bmatrix} \mapsto \begin{bmatrix} (w^2+z^2)^{1/2} \\ (w^2+z^2)^{-3/2} \bigl( w^3 x + 3 w^2 zy - 3 w z^2 x - z^3 y \bigr) \\ (w^2+z^2)^{-3/2} \bigl( w^3 y - 3 w^2 zx - 3 w z^2 y + z^3 x \bigr) \\ 0 \end{bmatrix} .
\end{equation}
In the polar form, it reduces to
\begin{equation}
q_{\textrm{rel}} = \cos\frac{\beta}{2} + \bigl( {\bf i} \cos\alpha + {\bf j}\sin\alpha \bigr) \sin\frac{\beta}{2} = \cos\frac{\beta}{2} + {\bf m}\sin\frac{\beta}{2}  ,
\end{equation}
which we use to extract $\alpha$ and $\beta$ information for each loop. The vector part uniquely determines the local profile and is used for plotting in Fig.~3 in the main text.

In addition to rotations around the tangent, reversing the sign of the tangent also results in the same profile. In parameterisation \eqref{eqS:n_phi}, this corresponds to reversing the parameter direction $\phi\mapsto -\phi$ and rotation of the director by $\pi$ around the ${\bf i}$ axis, resulting in the similarity transformation 
\begin{equation}
  q_{\textrm{rel}}\mapsto -{\bf i}q_{\textrm{rel}}{\bf i},
\end{equation}
which reflects the vector ${\bf m}$ over the ${\bf i}$ axis, inducing equivalency of states $(\alpha,\beta) \sim (-\alpha,\beta)$. The statistical analysis in Fig.~3(b) shows this symmetry, although on each particular defect loop (Fig.~3(a)), loop reversal mirrors the entire trajectory as a whole, so the profile variation is always continuous. There is no equivalent discrete gauge symmetry that would make the diagrams symmetric left-to-right, $(\alpha,\beta)$ states are not physically equivalent to $(\pi+\alpha,\beta)$; instead, reflection of ${\bf m}$ over ${\bf j}$ axis corresponds to inversion, reversing the handedness of the profile. This symmetry is bound to the physical chirality of the analysed active director texture; the material itself need not be chiral, an achiral material can also assume a chiral flow state (as is the case with active nematic droplets at lower activity \cite{copar2019S}).

\end{document}